\documentclass[10pt,conference]{IEEEtran}

\usepackage{threeparttable}
\usepackage[normalem]{ulem}
\usepackage{algpseudocode}
\usepackage{graphicx}
\usepackage{ifthen}
\usepackage[linesnumbered,ruled,vlined]{algorithm2e}
\usepackage{adjustbox}

\usepackage{amsmath}
\usepackage{tcolorbox}
\usepackage{varwidth}
\usepackage{mdframed}
\usepackage{multirow}
\usepackage{array}
\usepackage{tikz}
\usepackage{listings}
\usepackage{xcolor}
\usepackage{balance}
\usepackage{color}
\usepackage{fancybox}
\usepackage{cellspace}
\usepackage{comment}
\usepackage{diagbox}
\usepackage{url}
\usepackage{xspace}
\usepackage{amssymb}
\usepackage[caption=false]{subfig}
\usepackage{epsfig}
\usepackage[english]{babel}
\usepackage[numbers,sort&compress]{natbib}
\usepackage[T1]{fontenc}
\usepackage{titlesec}
\usepackage{titletoc}
\usepackage{framed}
\usepackage{tcolorbox}
\usepackage{fontawesome} 
\usepackage[hidelinks]{hyperref}
\usepackage[utf8]{inputenc}
\usepackage{pgfplots}
\usepackage{colortbl}
\usepackage{tikz}
\usepackage{etoolbox}

\usepackage[font=small,labelfont=bf]{caption}

\pgfplotsset{compat=1.17}
\usetikzlibrary{matrix}

\newtcolorbox{findingbox}[1][]{ #1}

\lstdefinestyle{bnf-style}{
    basicstyle=\ttfamily,
    keywordstyle=\color{blue},
    columns=fullflexible,
    keepspaces=true,
    morekeywords={::=, |},
    literate={->}{$\rightarrow$}{2}
            {epsilon}{$\varepsilon$}{1}
            {langle}{$\langle$}{1}
            {rangle}{$\rangle$}{1}
}

\newcommand{\am}[1]{\textcolor{red}{{\it [Comment: #1]}}}

\newcommand{\emna}[1]{\textcolor{purple}{{\it [Comment: #1]}}}

\title{Refactoring for Dockerfile Quality: A Dive into Developer Practices and Automation Potential}
\author{
    \begin{tabular}{c @{\hskip 4cm} c}
        Emna Ksontini & Meriem Mastouri \\
        {emna@umich.edu} & {meriemm@umich.edu} \\
        University of Michigan - Flint & University of Michigan - Flint \\
        USA & USA \\
        \\
        Rania Khalsi & Wael Kessentini \\
        {rkhalsi@umich.edu} & {wkessent@depaul.edu} \\
        University of Michigan - Flint & DePaul University \\
        USA & USA \\
    \end{tabular}
}

\begin{document}

\maketitle

\begin{abstract}

Docker, the industry standard for packaging and deploying applications, leverages Infrastructure as Code (IaC) principles to facilitate the creation of images through Dockerfiles. However, maintaining Dockerfiles presents significant challenges. Refactoring, in particular, is often a manual and complex process.

This paper explores the utility and practicality of automating Dockerfile refactoring using 600 Dockerfiles from 358 open-source projects. Our study reveals that Dockerfile image size and build duration tend to increase as projects evolve, with developers often postponing refactoring efforts until later stages in the development cycle. This trend motivates the automation of refactoring. To achieve this, we leverage In Context Learning (ICL) along with a score-based demonstration selection strategy. Our approach leads to an average reduction of 32\% in image size and a 6\% decrease in build duration, with improvements in understandability and maintainability observed in 77\% and 91\% of cases, respectively. 
Additionally, our analysis shows that automated refactoring reduces Dockerfile image size by 2x compared to manual refactoring and 10x compared to smell-fixing tools like PARFUM.

This work establishes a foundation for automating Dockerfile refactoring, indicating that such automation could become a standard practice within CI/CD pipelines to enhance Dockerfile quality throughout every step of the software development lifecycle.

\end{abstract}
\begin{IEEEkeywords}
    Refactoring, In-Context Learning, Software Engineering, LLM, Docker Refactoring, IaC.
\end{IEEEkeywords}
\section{Introduction}
\label{introduction}

Docker has emerged as the de facto industry standard, offering a platform for creating, deploying, and managing containers \cite{bullington2020docker}. The contents of a Docker container are defined by Dockerfile declarations, which specify the instructions and their execution order. Often found in source code repositories, these files facilitate the building of images that can then be run as containers, thereby bringing the hosted software to life in its execution environment \cite{miell2019docker}.

Despite being the backbone of containerization, Dockerfiles are often plagued with problems such as smells \cite{durieux2024empirical}. In this context, several works have been proposed to detect and fix them \cite{yiwen2020characterizing,bui2023dockercleaner,durieux2023parfum}. However, most solutions focus on bash commands, which are typically embedded within specific instructions in Dockerfiles, mainly the RUN instruction. Common bash smells include the installation of unnecessary packages and missing version pinning \cite{yiwen2020characterizing,rosa2024not}. While bash-based solutions are useful, they fall short in addressing the broader structure of Dockerfiles, which are not limited to the RUN instruction. Instead, Dockerfiles include various instructions and stages that together form the final image.

Developers face challenges beyond managing shell scripts in Dockerfiles, such as reducing technical debt, as defined in \cite{ksontini2021refactorings}, which affects all instructions and stages, compromising overall quality. For instance, a key challenge is maintaining small \textbf{image sizes}. Each instruction within a Dockerfile adds a layer to the final image, potentially leading to bloated containers if not managed carefully. Choosing the appropriate base image and effectively utilizing multi-stage builds can mitigate this issue, but achieving the right balance requires considerable expertise and effort \cite{straesser2023empirical, durieux2024empirical}.
\textbf{Maintaining} Dockerfiles is another challenge, with common bugs and unexpected behaviors often arising, especially when the base image tag is not specified correctly. Additionally, some developers, unaware of the availability of official and trusted images, manually add dependencies, which can lead to inconsistencies and complicate future updates \cite{azuma2022empirical}. \textbf{Build duration} is influenced by several factors, including the order of instructions, the complexity of multi-step commands, and the number of instructions. Poorly optimized sequences and complex instructions can lead to high build durations and incur high costs, complicating the workflow \cite{wu2022understanding}. As a project progresses through its development stages, the content of the Dockerfile may be revised many times \cite{9359307}, which can affect \textbf{understandability} and complicate subsequent updates.

Current refactoring tools designed for languages like Java, JavaScript, C, and Python \cite{golubev2021one} are inadequate for Dockerfiles due to the distinct nature of IaC and their declarative syntax. Furthermore, Dockerfiles are used to containerize applications across a wide array of programming languages (e.g., Java, PHP, Go) and domains (e.g., machine learning, web development), which requires refactoring efforts that extend beyond mere structural changes. Effective refactoring often demands domain expertise in selecting appropriate base images, specifying correct tags, and adjusting dependencies. This complicates the application of search-based techniques in this context. At the same time, using deep learning (DL) approaches poses major challenges due to its data-intensive nature. Even with access to the largest Dockerfile collection available in the literature, DL models fell short in capturing the intricate patterns and variability unique to Dockerfiles \cite{rosa2023automatically}.

To address the data shortage problem, we propose leveraging Large Language Models (LLMs) \cite{openai2023gpt4}, which have recently emerged as powerful tools for various software engineering tasks \cite{hou2023large}. These models are pre-trained on vast datasets, including billions of code repositories, 
enabling them to capture substantial domain knowledge. We hypothesize that this exposure allows LLMs to generate Dockerfile refactorings that closely align with common practices and are likely to be adopted by developers.

To optimize LLM performance for this purpose, we employ ICL. Recent studies show that ICL, as opposed to straightforward prompts, effectively leverages the domain knowledge embedded within LLMs by aligning input formats with those used in pre-training \cite{nashid2023retrieval,rubin2021learning}. 
This strategy aligns with recent industry trends: Docker recently launched labs with a generative AI assistant (based on the GPT-4 model) to aid Dockerfile creation, and they reported that prompts lacking best practices led to lower-quality Dockerfiles \cite{dockerlabs2024}.

In this paper, we introduce a novel approach that harnesses the capabilities of ICL to automate Dockerfile refactoring. Our methodology involves four key components: (1) an analysis of how developers currently perform Dockerfile refactoring within open-source projects, aiming to assess the utility and potential benefits of automating these tasks; (2) an evaluation of ICL effectiveness in automating refactoring, using a novel score-based selection method for demonstration examples, with a focus on optimizing image size, build duration, understandability, and maintainability; (3) a comparative analysis of automated refactoring, manual refactoring, and smell-fixing tools with respect to their effectiveness in enhancing Dockerfile quality, and (4) a thorough examination of the reasons behind build failures caused by refactoring.

To the best of our knowledge, this is the first study on Dockerfile refactoring automation, revealing several findings:
\begin{itemize}
\item Developers predominantly engage in Dockerfile refactoring during the mid to late stages of project development, often addressing accumulated technical debt.
\end{itemize}

\begin{itemize}
\item In general, the effectiveness of ICL in automating Dockerfile refactoring significantly improves with an increasing number of demonstrations.
\end{itemize}

\begin{itemize}
\item Automated refactoring outperforms both manual refactoring and smell repair tools, achieving greater reductions in Docker image size and build duration, while also improving the understandability and maintainability.
\end{itemize}

\begin{itemize}
\item The primary causes of build failures in both automated and manual Dockerfile refactorings are related to build context errors and dependency management.
\end{itemize}

\textbf{Replication Package.} All materials are available in ~\cite{ICSE25_replication}.
\section{Related Work}
\label{sec:relatedwork}

\subsection{Docker Quality Issues}

Docker's documentation offers best practices \cite{bestpractice} for addressing Dockerfile smells. However, developers' adherence to these guidelines is inconsistent. These smells, akin to traditional configuration code smells \cite{ushar2016does}, indicate design flaws that, although not bugs, negatively affect Docker images. 

In \cite{yiwen2020characterizing}, authors have categorized Dockerfile smells into two primary types: \textit{DL}-smells, which breach Docker's official best practices, and \textit{SC}-smells, which violate basic shell script conventions. Their empirical study revealed that most open-source projects exhibited Dockerfile smells, with \textit{DL}-smells occurring more frequently than \textit{SC}-smells. Similarly, Durieux et al. \cite{durieux2024empirical} have investigated how shell smells in Dockerfiles increase image size, affecting distribution efficiency and scalability.

Henkel et al. \cite{jordan2020learning} have developed Binnacle to mine Bash-related smells from over 178,000 Dockerfiles. Xu et al. \cite{iwei2019dockerfile} have introduced \textit{TF}-smells, showing how the risky practice of leaving temporary files in Docker images during the build process can inflate image size and complicate distribution.

 Beyond detection, some studies have explored the automation of Dockerfile smell repairs. For example, \textit{PARFUM} \cite{durieux2023parfum} detects and automatically repairs Dockerfile smells, evaluating repair effectiveness in terms of build failure and image size. Rosa et al.\cite{rosa2024not} have evaluated Hadolint writing practices by experts and ranked 26 additional Dockerfile smells, offering guidance for high-quality Dockerfile writing. 
 
 Other studies, such as \cite{wu2020exploring}, used linear regression to analyze Dockerfile quality and project characteristics, revealing statistical correlations. Cito et al. \cite{Cito2017MSR} found that missing version pinning smell in Dockerfiles often causes build issues, leading to non-reproducible builds. In \cite{henkel2021shipwright}, a human-in-the-loop system using a modified BERT model has been developed for repairing Dockerfiles, showing potential for automating repairs and boosting build success. 
 
While most studies focus on addressing Bash smells within the RUN instruction, Dockerfiles encompass a broader range of instructions and stages that can impact overall quality. Addressing these broader quality issues is crucial, as they often signal technical debt in Docker projects. Such debts can manifest as excessively large images, which in turn lead to slower deployment times and increased resource consumption. Refactoring these elements is essential to improve performance and efficiency in Docker-based environments \cite{ksontini2021refactorings}.

In \cite{ksontini2024drminer}, DRMiner has been introduced for identifying and analyzing Dockerfile refactorings, providing semantics-aware static analysis. Although this tool indicates future potential for automated Dockerfile refactoring, to the best of our knowledge, no tool currently achieves full automation. Similarly, ~\cite{rosa2022assessing} have suggested methods to improve Dockerfile quality and automate the generation of high-quality Dockerfiles. These studies shift focus from detecting shell smells to addressing border issues through potential automation, enhancing Dockerfile quality.

In this paper, we fully automate the refactoring of Dockerfiles to reduce technical debt and improve their quality.

\subsection{Refactoring Recommendation and Automation}

Several research efforts have aimed to assist or automate code refactoring using various techniques. Despite the availability of automated tools, studies reveal that refactoring is often performed manually due to issues like tool integration and lack of support for real-world scenarios \cite{eilertsen2020refactoring}.

Search-based techniques \cite{harman2012search} treat software refactoring as an optimization problem aimed at enhancing system design quality using software metrics. Ouni et al. \cite{ouni2012search} developed a multi-objective formulation to address code smells, while Alizadeh et al. \cite{alizadeh2018interactive} employed NSGA-II for dynamically and interactively recommending refactoring solutions.

ML techniques have also been widely used for recommending refactorings. Nyamawe et al. \cite{nyamawe2020feature} utilized classifiers like LR, CNN, and SVM to suggest refactorings based on historical data and detected code smells. Sagar et al. \cite{Sagar2021} created a dataset of five refactoring types using 800 open-source Java projects, applying RF, SVM, and LR for identification.

Recent studies highlight in-context learning's potential to enhance LLM performance in software engineering tasks \cite{incontext2024}. Zhang et al. \cite{zhang2024refactoring} showed LLMs have promising bug-fixing abilities. Madaan et al. \cite{madaan2023selfrefine} used Codex, to improve code readability, while Shirafuji et al.\cite{shirafuji2023refactoring} demonstrated \textit{GPT-3.5}'s effectiveness in refactoring Python programs, significantly reducing complexity and code lines.
 
Despite these advancements, there is a notable gap in tools specifically designed for IaC, particularly for Docker. Srivatsa et al. \cite{srivatsa2024survey} emphasized the potential of LLMs in generating IaC configurations, suggesting automation can make these processes more accessible.

To our knowledge, \textbf{no} studies have automated Dockerfile refactoring or applied ICL to improve Docker quality. This research investigates automated refactoring, comparing its effectiveness to manual refactoring and smell-fixing tools.

\section{STUDY DESIGN}
\label{sec:04-Approach}

\subsection{Research Questions}

\begin{figure*}[!h]
    \centering
    \includegraphics[scale=0.3]{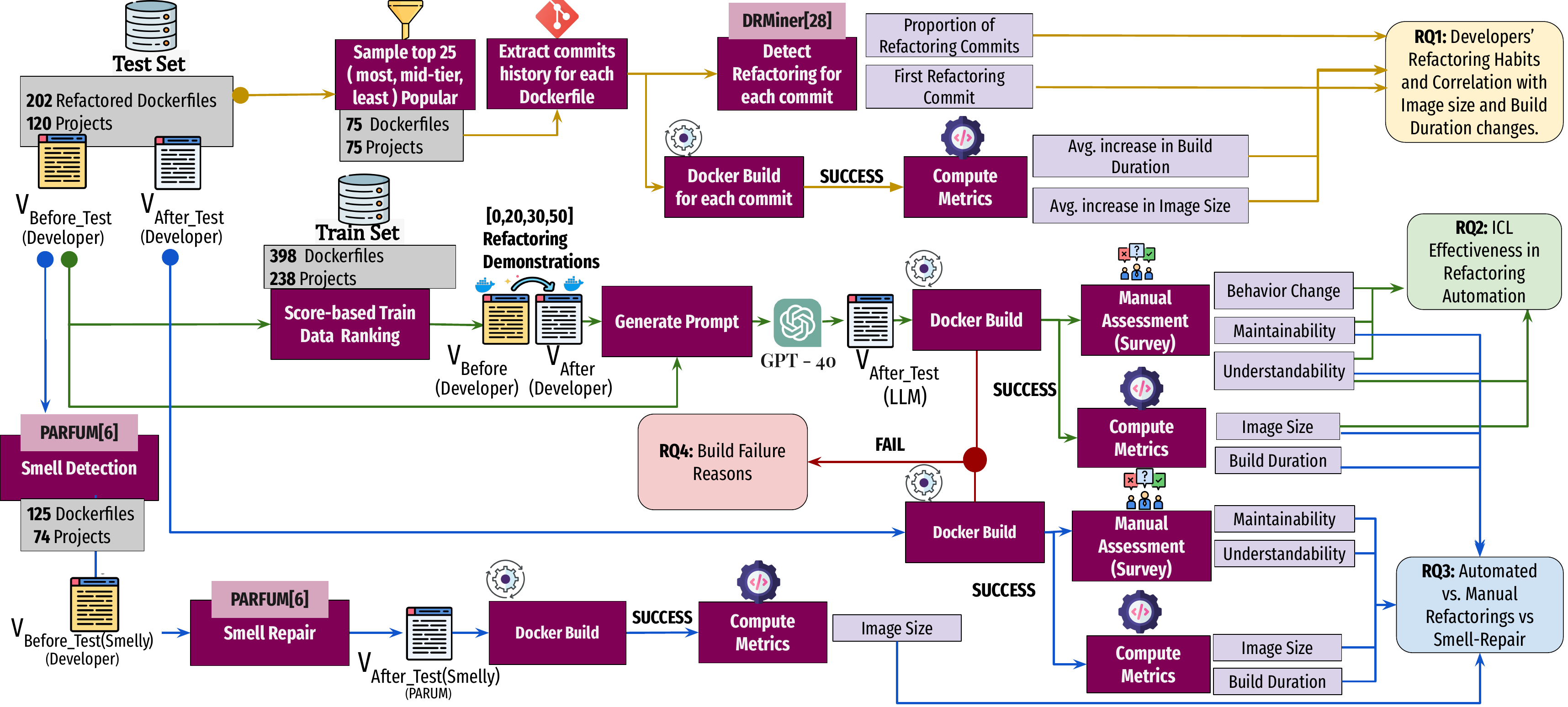}
    \caption{Approach overview and addressed RQs (RQ1: yellow arrows, RQ2: green arrows, RQ3: blue arrows, and RQ4: red arrows) }
    \label{fig:apprach}
\end{figure*}

Our study investigates the practicality and efficiency of automating Dockerfile refactoring. We address the following research questions, summarized in \autoref{fig:apprach}.

\textbf{RQ1: To what extent do developers' refactoring habits correlate with changes in Docker image size and build duration?}

Getting insights into how Dockerfiles evolve in open-source projects is crucial for understanding the benefits of automating their refactoring. Key aspects include identifying when developers refactor Dockerfiles, the frequency of these refactorings, and the impact of Dockerfile changes on performance metrics such as build duration and image size. Analyzing these elements can highlight the time-saving opportunities and performance improvements that automation can offer.

To address RQ1, we perform a quantitative analysis of 75 Dockerfiles from 75 open-source GitHub projects, illustrated in \autoref{fig:apprach} (yellow arrows). For the commit history of each Dockerfile, we identify instances of refactoring using DRMiner \cite{ksontini2024drminer}. We then map out the lifecycle stages during which these refactoring actions occurred. Given the varying number of commits and lifetimes of the studied projects, each project's lifecycle is segmented into ten equal phases: the first 10\% of the total commits, the next 10\%, and so on. In this context, we exclude projects with less than ten commits.

This allows for determining when developers typically start refactoring and during which stage it happens most frequently. 

Beyond identifying refactoring instances, we automatically build each Dockerfile for every commit in its history. If successful, we measure the build duration and image size to study their average increase during the Dockerfile's evolution.


\textbf{RQ2: How effective is ICL in automating Dockerfile refactoring using zero-shot, one-shot, and few-shot learning approaches?}

This research question evaluates the effectiveness and reliability of ICL for automating Dockerfile refactoring, focusing on how zero-shot, one-shot, and few-shot approaches impact model performance.

To answer \textbf{RQ2}, we utilize a dataset comprising 202 refactored Dockerfiles from 120 unique open-source projects. As illustrated in \autoref{fig:apprach} (green arrows), we start by prompting the LLM with the pre-refactoring version of each Dockerfile ($V_{Before\_Test}$), using a zero-shot approach. Next, we examine the effectiveness of one-shot and few-shot learning by including one or more refactoring demonstrations in the prompt. 
To select these examples, We develop a customized selection strategy that employs score-based ranking to identify the most relevant refactoring demonstrations from a training set of 398 Dockerfile refactoring commits across 238 projects. Moreover, during few-shot learning, we evaluate the impact of increasing the number of demonstrations on the model's performance.

During these experiments, each generated Dockerfile refactoring undergoes a behavioral assessment to ensure its functional behavior remains unchanged. Finally, we evaluate effectiveness using several key metrics, including build failure rates, maintainability, understandability, image size, and build duration.


\textbf{RQ3: How effective is automated Dockerfile refactoring in improving quality, compared to manual refactoring and smell-detection tools?}

In this RQ, we aim to determine how well Dockerfile refactorings automation compare to those performed by developers. This evaluation will offer insights into the practicality of automated refactoring in real-world software development. 

Moreover, we found that
existing techniques, such as PARFUM \cite{durieux2023parfum}, an automated tool for repairing Dockerfile bash-related smells, have demonstrated measurable improvements in Dockerfile quality, notably in reducing image size \cite{durieux2024empirical}. We aim to contrast the efficacy of improvements achieved through automated refactoring with those driven purely by smell fixing. We compare with PARFUM since it is a recent work offering automated repairs or Dockerfile smells. In contrast, other techniques, such as Shipwright \cite{henkel2021shipwright}, Hadolint \cite{hadolint95:online}, and Binnacle \cite{jordan2020learning}, focus on detecting error patterns or smells and require manual intervention for their operation.

To answer \textbf{RQ3}, we use the results from the best-performing ICL approach identified in RQ2. We first compare automatically refactored Dockerfiles, denoted ($V_{\text{After}\_\text{Test}}(\text{LLM)}$), with their human-refactored versions, denoted ($V_{\text{After}\_\text{Test}}(\text{Developer)}$), as illustrated in \autoref{fig:apprach} (blue arrows). This comparison is based on the same effectiveness aspects used in RQ2: build failure, image size, build duration, understandability, and maintainability. We then contrast the results of automated refactoring with those achieved solely through smell correction, specifically noted as ($V_{\text{After}\_\text{Test(smelly)}}(\text{PARFUM)}$), focusing on image size as a core outcome measure. For a fair comparison with PARFUM, we restricted this analysis to Dockerfiles identified as ``smelly", representing 125 instances from our test set.


\textbf{RQ4:What causes Build Failures after refactoring, and how do these reasons differ between automated and manual approaches?}

In \textbf{RQ2} and \textbf{RQ3}, it was observed that while the ICL achieved commendable improvements in image size, build duration, understandability, and maintainability, there was a considerable rate of build failures in the generated refactored Dockerfiles. Interestingly, we observed a similar rate of build failures in the Dockerfiles refactored by human developers. In this research question, we aim to understand the reasons behind these build failures. By uncovering these reasons, the aim is to identify further hidden factors related to Dockerfile refactoring practices and improve the automation process.

To answer \textbf{RQ4}, we conduct a qualitative analysis of build failures caused by both ICL and human developers over a set of 182 Dockerfile refactoring cases. The steps for this analysis are illustrated in \autoref{fig:apprach} (red arrows).
\subsection{Refactored Dockerfiles Data Collection}
To construct our dataset, we gather refactored Dockerfiles from the commit history of Docker projects, focusing on commits created primarily for refactoring. The parent versions of these Dockerfiles are assumed to have quality problems, as evidenced by the developers' choice to refactor them. This makes them relevant candidates for assessing the effectiveness of automated refactoring processes. The detailed extraction and filtering procedures figure is available in \cite{ICSE25_replication}.

To begin with, we use the Google BigQuery GitHub Public Dataset \cite{ICSE25_replicATION_dataset}, released in 2016, to source our data. We focus on the most recent snapshot available as of March 10, 2024, and target commit messages containing terms referring to self-affirmed refactoring patterns such as ``\textit{refactor.*}'', ``\textit{fix.*}'', ``\textit{improve.*}'', following findings from AlOmar et al. \cite{alomar2019can}. In addition, we ensure that these commits mention the word ``\textit{Dockerfile}''. This query yields 3,046 Dockerfiles from  1634 commits across 1,293 projects, representing potential pre- and post-refactoring versions of a Dockerfile ($\text{V}_{\text{before}}$ and $\text{V}_{\text{after}}$). 

Next, we perform an automated build, and Dockerfiles that fail to build in their pre-refactored state are discarded, leaving us with 650 successful Dockerfile pairs from 381 projects. This ensures that we are working with functional Dockerfiles before applying automation.


We process the dataset further using DRMiner \cite{ksontini2024drminer}, a tool that detects specific refactoring actions performed on Dockerfiles. While DRMiner is designed to identify 12 refactoring types, the ``\textit{Move Stage}'' and ``\textit{Extract Run Instruction}'' types are only present in two Dockerfiles in our dataset. Therefore, we decide to omit these two types and focus on the remaining 10 refactoring types. After this step, we are left with 641 Dockerfiles from 375 projects.

To ensure that the dataset only includes refactoring changes, three authors examined the Dockerfile pairs and excluded files with functional behavior changes. Functional behavior refers to Dockerfile elements that affect the application's runtime behavior, namely application files (\textit{COPY}/\textit{ADD}) and startup commands (\textit{ENTERPOINT}/\textit{CMD}). Following this step, we retained 600 Dockerfiles from 358 projects.

At this stage, we split the dataset into training (75\%) and test (25\%) sets using stratified sampling to ensure balanced representation of refactoring types. To avoid having Dockerfiles from the same projects in both sets, we moved about 2\% (12 Dockerfiles) to the test set, resulting in 27\% (162 Dockerfiles) in the test set and 73\% (438 Dockerfiles) in the training set.
Finally, we verify that all Dockerfiles in the training set were successfully built post-refactoring. This additional verification step resulted in a further 7\% (40 Dockerfiles) of the data being removed from the training set and moved to the test set due to build failures. The final dataset distribution comprises 398 Dockerfile revisions (approximately 66\%) from 238 unique projects in the \textbf{training set} and 202 Dockerfile revisions from 120 unique projects (approximately 34\%) in the \textbf{test set}.
\subsection{The Prompt Template for Refactoring Automation}

Our prompt is defined as: $P = \{ N +RD +V_{\text{Before}\_\text{Test}} \}$ where $N$ is a natural language template specifying the task description and the definition of all possible refactoring actions \cite{ksontini2021refactorings}. $RD = \{V_{\text{Before}_i}, V_{\text{After}_i}, R_i \}_{i=1}^n$ is a set of refactoring demonstrations composed of the input Dockerfile $V_{\text{Before}_i}$, the desired output Dockerfile $V_{\text{After}_i}$, and the refactoring actions applied $(R_i)$. $V_{\text{Before}\_\text{Test}}$ is a Dockerfile for testing. Specifically, if $n = 0$, indicating no refactoring demonstration, the setting is known as \textit{zero-shot learning}; if $n = 1$, indicating only one refactoring demonstration, the setting is known as \textit{one-shot learning}; and \textit{few-shot learning} applies when there are multiple refactoring demonstrations. In addition, there is a constraint that $\text{size}(P) \leq \text{context\_window}$, which means that the prompt should fit within the token limit of the language model.  
A figure detailing the prompt can be found in \cite{ICSE25_replication}.

\subsection{Refactoring Demonstration Retrieval}
\label{subsec:demo}

Existing research demonstrates that ICL achieves better performance in code intelligence tasks with instance-specific demonstrations tailored to test inputs rather than task-level ones \cite{incontext2024,min2022rethinking,nashid2023retrieval}. 
Drawing on this finding, in RQ2 we develop a score-based demonstration selection strategy based on technical debts defined in \cite{ksontini2021refactorings}. 
As illustrated in \autoref{fig:apprach}. Given a test Dockerfile $V_{\text{Before}\_\text{Test}}$, we identify the most relevant dockerfile refactoring demonstrations \( RD_i \) using the following formula: 

\vspace{-2mm}
    \begin{equation}
\setlength{\arraycolsep}{1pt}
\renewcommand{\arraystretch}{0.8}
\scalebox{0.9}{ 
  \fbox{
    \begin{minipage}{0.8\linewidth}
      \small
      \begin{align*}
      \text{score}(RD_i) = & \, 0.2 \times \text{Textual\_similarity}(V_{\text{Before}_i}, V_{\text{Before}\_\text{Test}}) + \\
                           & \, 0.2 \times \text{understandability\_score}(V_{\text{Before}_i},V_{\text{After}_i}) + \\
                           & \, 0.2 \times \text{maintainability\_score}(V_{\text{Before}_i}, V_{\text{After}_i}) + \\
                           & \, 0.2 \times \text{image\_size\_score}(V_{\text{Before}_i}, V_{\text{After}_i}) + \\
                           & \, 0.2 \times \text{build\_duration\_score}(V_{\text{Before}_i},V_{\text{After}_i})
      \end{align*}
    \end{minipage}
  }}
\end{equation}
The components of the score are defined as follows:

\begin{itemize}
    \item \textbf{Textual\_similarity}: Calculated using BM-25 \cite{robertson1995okapi}.
    \item \textbf{Understandability (or Maintainability) score}: Assigned a value of 1 if the understandability (or maintainability) of \( V_{\text{After}_i} \) is greater than \( V_{\text{Before}_i} \), 0 if they are the same, and -1 if it is worse.
    \item \textbf{Build duration (or image size) score}: Calculated as  

\( 1 - \frac{\text{build\_duration}(\text{or  image\_size) }(V_{\text{After}_i})}{\text{build\_duration} ( \text{or  image\_size)}(V_{\text{Before}_i})} \), quantifying the improvement in build duration  (or image size).
\end{itemize}

We use an automated script to compute the image size (MB) and build duration (s). To ensure consistency, we measure the build duration over three runs and use the average. For maintainability and understandability, we employ a manual assessment performed by three authors of the paper. Each Dockerfile was labeled three times, with assessments deemed valid if at least two evaluators agreed. 

For each query, we compute the score over the entire corpus of demonstrations, rank the train corpus based on this score, and select the top $n$ based on the number of shots. For instance, we select the top example for one-shot, the top 20 examples for 20-shot, and so on. When adding the demonstrations to the prompt, we order them in ascending order, meaning the one with the highest score will be closest to the query Dockerfile. This ordering follows the findings by Gao et al. \cite{incontext2024}.

This scoring strategy returns demonstrations that closely match the query Dockerfile and exhibit the greatest overall quality improvements. By focusing on content similarity, we ensure that the model understands the specific needs of different Dockerfiles in terms of applications and commands. Moreover, prioritizing quality improvements enables the model to learn optimal practices from the best examples.

\begin{figure*}[h!]
    \centering
    \begin{minipage}[t]{0.32\textwidth}
        \centering
        \includegraphics[width=\textwidth]{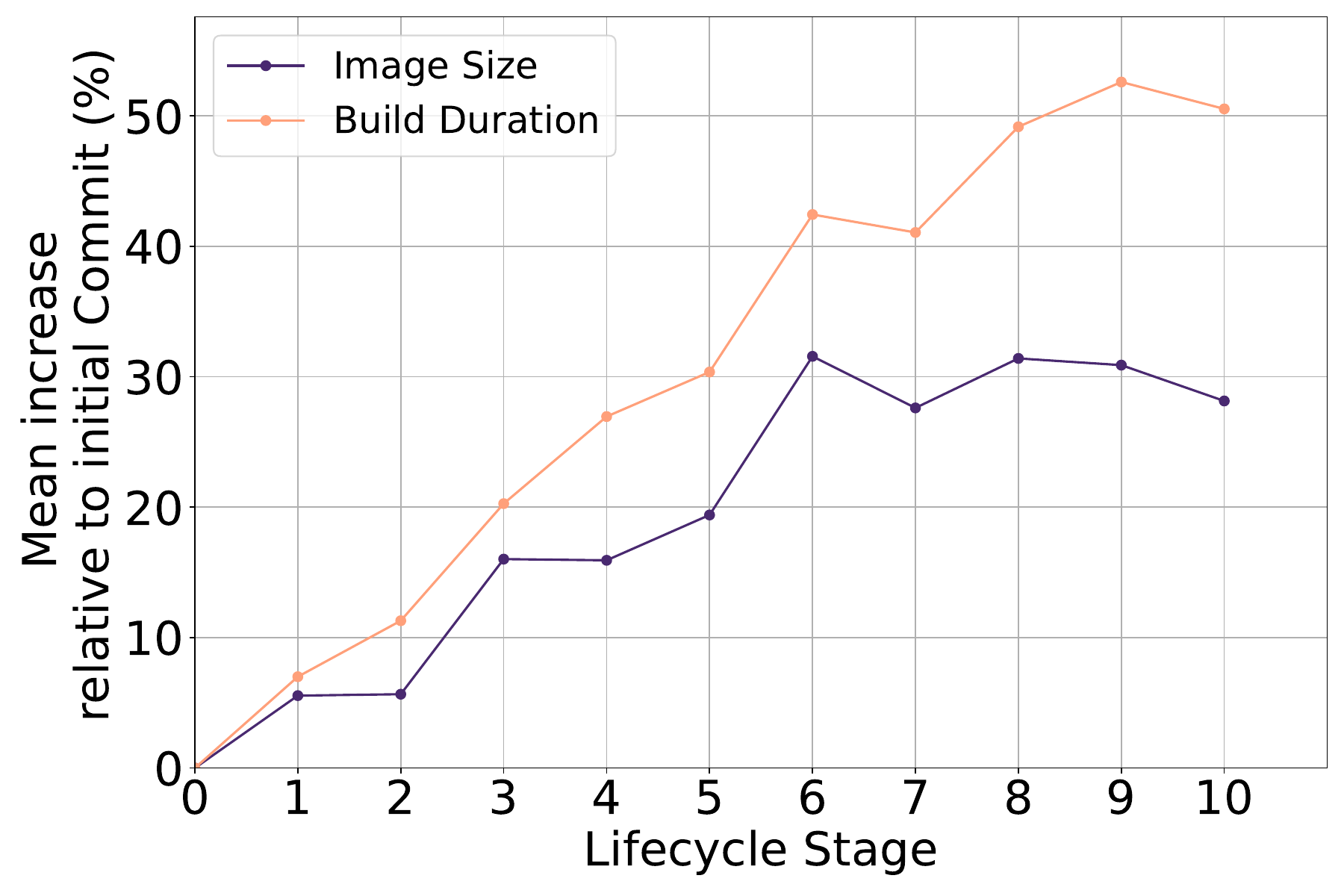}
        \captionsetup{width=1\textwidth} 
        \caption{Mean Image Size \& Build Duration Increase Over Project Lifecycle}
        \label{fig:build_image}
    \end{minipage}
    \hfill
    \begin{minipage}[t]{0.32\textwidth}
        \centering
        \includegraphics[width=\textwidth]{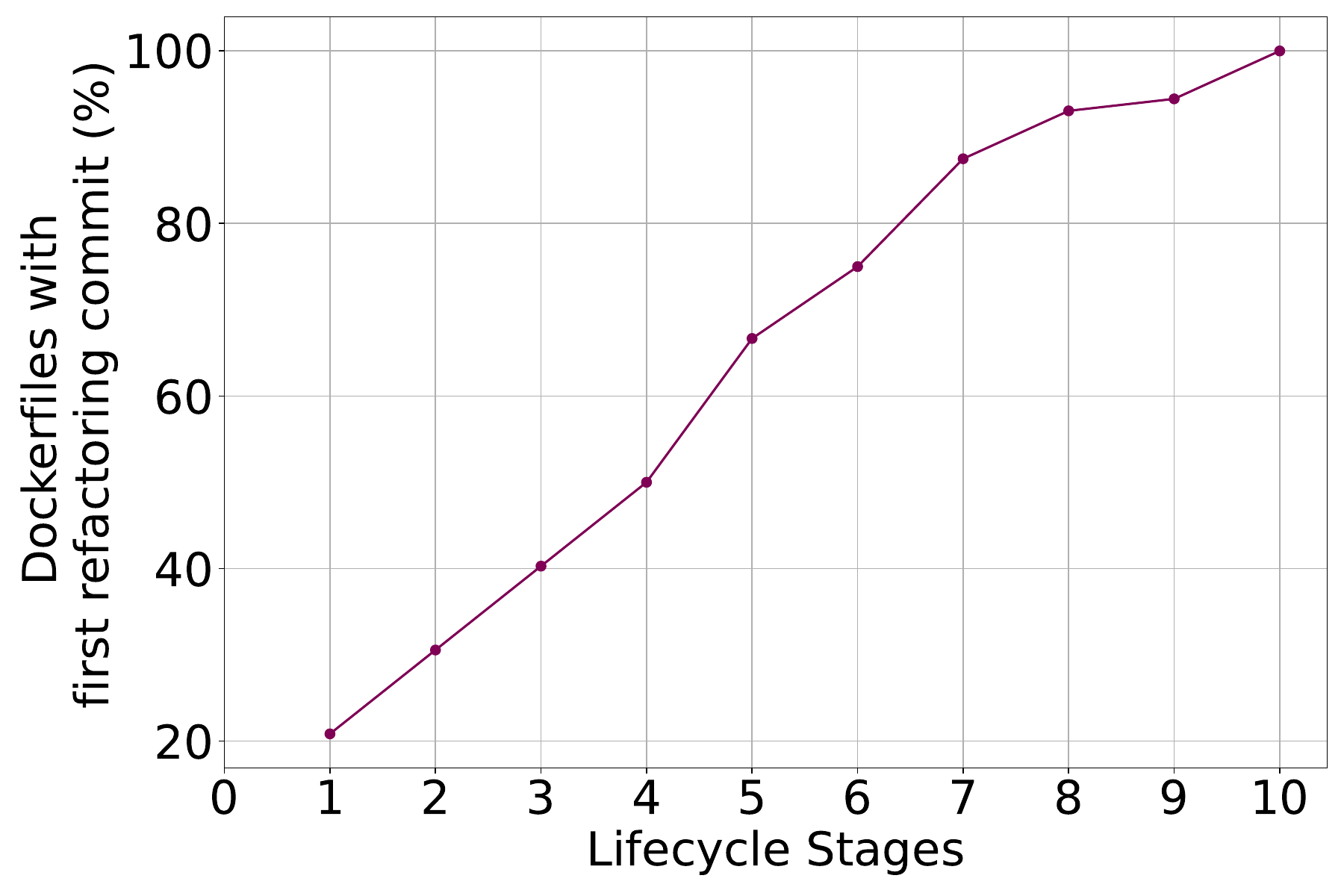}
        \captionsetup{width=1\textwidth}
        \caption{Cumulative percentage of Dockerfiles with first refactoring commit.} 
        \label{fig:per_ref}
    \end{minipage}
    \hfill
    \begin{minipage}[t]{0.32\textwidth}
        \centering
        \includegraphics[width=\textwidth]{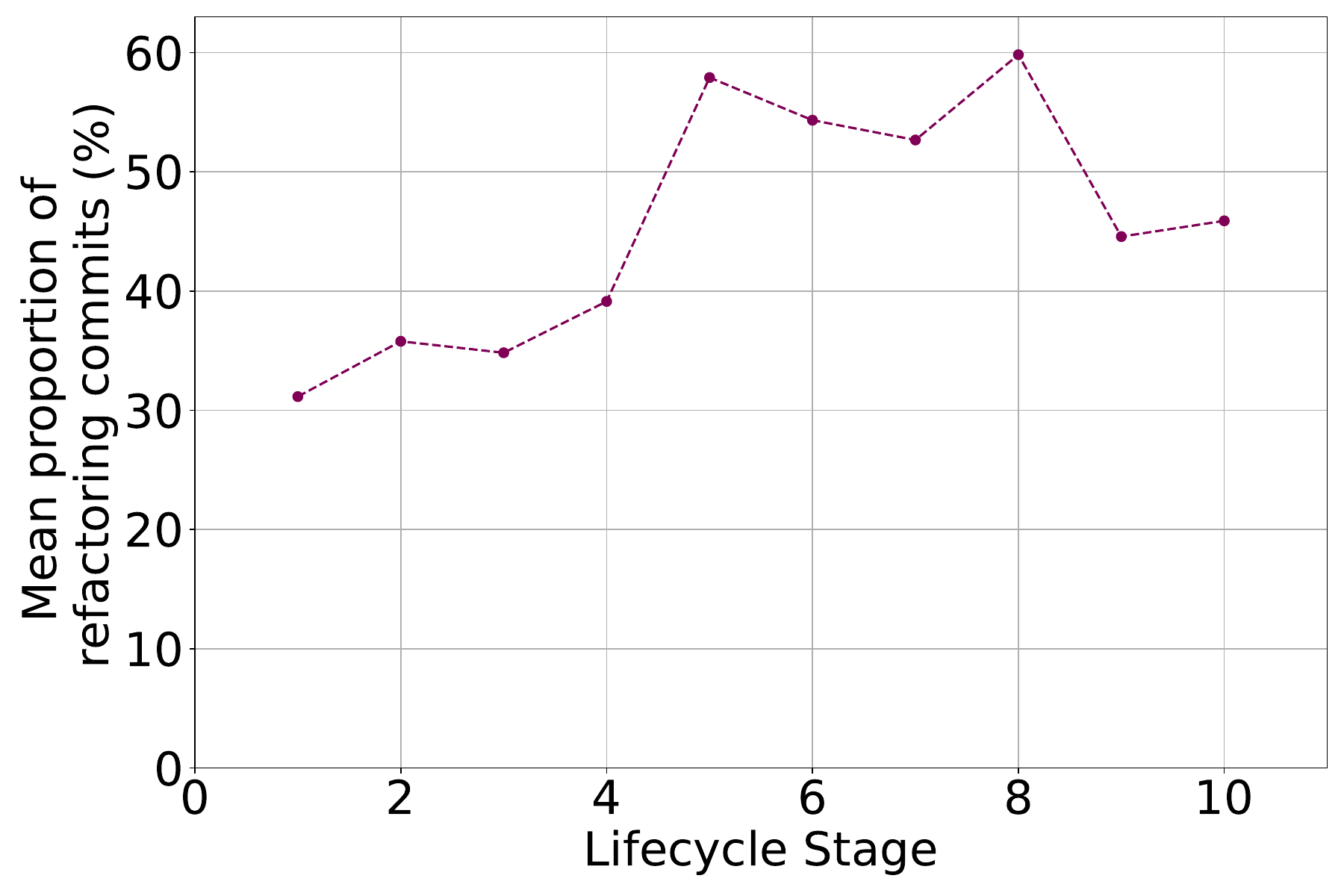}
        \captionsetup{width=1 \textwidth}
        \caption{Mean proportion of refactoring commits across project lifecycles.}
        \label{fig:propo_ref}
    \end{minipage}

\end{figure*}

\subsection{Evaluation Metrics}

To assess the effectiveness of manual and automated refactorings in RQ2 and RQ3, we also rely on the technical debt outlined in \cite{ksontini2021refactorings}. This involves measuring the image size and build duration for each Dockerfile version: original, manually refactored, and automatically refactored. For maintainability and understandability, we use the scoring system detailed in \autoref{subsec:demo}.

Six industry developers, each with about two years of Dockerfile experience, evaluated maintainability and understandability in an industry-sponsored collaboration. To ensure unbiased evaluations, developers were provided with pairs of Dockerfiles (original and refactored) without being informed which was the refactored version. They were asked to score the improvements in maintainability and understandability. For cases involving LLM-generated refactorings, developers were asked to report any changes in the Dockerfile's behavior, ensuring that any functional discrepancies were identified (manual refactoring cases had been pre-screened for behavioral changes during data collection).

Participants completed a pre-study survey on their programming experience and company roles. To minimize bias, they received a two-hour lecture on Docker quality assessment and refactoring, covering general and specific Dockerfile practices.

For result annotation, we used the Kappa cross-validation process. Each participant evaluated 300 Dockerfiles, with each pair assessed by three developers. An assessment was considered valid if at least two evaluators agreed, achieving an overall Cohen’s kappa of 0.84. Detailed information about the participants and evaluated Dockerfiles is available in our replication package \cite{ICSE25_replication}.

\subsection{Experiment Settings}

The rationale behind in-context learning is that LLMs, having been trained on vast corpora, absorb substantial domain knowledge, enabling them to generalize well to new tasks without the need for fine-tuning \cite{brown2020language}.

Among these LLMs, GPT-4 (Generative Pre-trained Transformer 4) has shown remarkable performance not only in conventional code-related tasks \cite{poldrack2023ai} but also in IaC scenarios \cite{namrud2024kubeplaybook}. Consequently, we have selected GPT-4o (an optimized variant of GPT-4) as the primary model for our experimental framework. Regarding the hyperparameters of the model's API, and in alignment with previous studies \cite{incontext2024,nashid2023retrieval}, we set the temperature to 0 to obtain deterministic outputs.


The context window of GPT-4o, which accommodates up to 128k tokens, imposes practical limits on the number of demonstrations we can use. Given that the largest Dockerfiles in our dataset are around 1200 tokens, and the combined size of the task description and refactoring action definitions is 434 tokens, each demonstration (involving two Dockerfiles plus refactoring actions) totals 2447 tokens. Based on these calculations, we determine that we can safely include up to 50 refactoring examples without exceeding the context window.

To explore different sizes of input examples, our experimental setup includes zero-shot, one-shot, and few-shot settings. For the few-shot settings, we incorporate 20, 30, and 50 refactoring demonstrations. We select 20 and 30 as intermediate steps to balance between the minimum (1-shot) and maximum (50-shot) feasible demonstrations, allowing us to observe the model's performance across this range.


All experiments were conducted on an 8-Core workstation equipped with an Intel(R) Xeon(R) CPU and 64 GB of RAM, operating on Ubuntu 18.04.


\section{Results}
\label{sec:05-Results}
\subsection{RQ1: Developers' refactoring habits and correlation with image size and build duration changes}
To address RQ1, we need to build each Dockerfile for every commit within the project. To ensure a balance between representativeness and computational feasibility, we selected a sample of Dockerfiles based on GitHub star counts. This sample includes 25 Dockerfiles each from high-, mid-, and low-popularity projects, representing a variety of languages and domains. In total, we selected 75 Dockerfiles from 75 projects. The demographics of these Dockerfiles are detailed in \cite{ICSE25_replication}.

\begin{table*}[h!]
\caption{Evaluation of Prompting Techniques and Developer Performance}
\begin{adjustbox}{max width=\textwidth}
\begin{tabular}{lc|cccc|cccc|cc|cc|}
\cline{3-14}
\multicolumn{2}{l|}{}                                                                                                                                                    & \multicolumn{4}{c|}{Image Size}                                                                                                                                                                                                                                                                                                                       & \multicolumn{4}{c|}{Build Duration}                                                                                                                                                                                                                                                                                                                    & \multicolumn{2}{c|}{Understandability}                                                                                                                       & \multicolumn{2}{c|}{Maintainability}                                                                                                                         \\ \hline
\multicolumn{1}{|l|}{\begin{tabular}[c]{@{}l@{}}Prompting\\ Techniques /\\ Developer\end{tabular}} & \begin{tabular}[c]{@{}c@{}}Build\\ Success\\ Rate (\%)\end{tabular}  & \multicolumn{1}{c|}{\begin{tabular}[c]{@{}c@{}}Improvement\\ Rate (\%)\end{tabular}}      & \multicolumn{1}{c|}{\begin{tabular}[c]{@{}c@{}}Deterioration\\ Rate (\%)\end{tabular}}   & \multicolumn{1}{c|}{\begin{tabular}[c]{@{}c@{}}Average\\ Reduction\end{tabular}}            & \begin{tabular}[c]{@{}c@{}}Total\\ Reduction\\ (GB)\end{tabular} & \multicolumn{1}{c|}{\begin{tabular}[c]{@{}c@{}}Improvement\\ Rate (\%)\end{tabular}}     & \multicolumn{1}{c|}{\begin{tabular}[c]{@{}c@{}}Deterioration\\ Rate (\%)\end{tabular}}   & \multicolumn{1}{c|}{\begin{tabular}[c]{@{}c@{}}Average\\ Reduction\end{tabular}}         & \begin{tabular}[c]{@{}c@{}}Total\\ Reduction\\ (minutes)\end{tabular} & \multicolumn{1}{c|}{\begin{tabular}[c]{@{}c@{}}Improvement\\ Rate (\%)\end{tabular}}    & \begin{tabular}[c]{@{}c@{}}Deterioration\\ Rate (\%)\end{tabular}  & \multicolumn{1}{c|}{\begin{tabular}[c]{@{}c@{}}Improvement\\ Rate (\%)\end{tabular}}    & \begin{tabular}[c]{@{}c@{}}Deterioration\\ Rate (\%)\end{tabular}  \\ \hline
\multicolumn{1}{|l|}{Zero-shot}                                                                   & \begin{tabular}[c]{@{}c@{}}38\%\\ (77/202)\end{tabular}           & \multicolumn{1}{c|}{\begin{tabular}[c]{@{}c@{}}81\%\\ (61/75)\end{tabular}}            & \multicolumn{1}{c|}{\begin{tabular}[c]{@{}c@{}}16\%\\ (12/75)\end{tabular}}           & \multicolumn{1}{c|}{\begin{tabular}[c]{@{}c@{}}129 MB\\ (18\%)\end{tabular}}          & 9 GB                                                          & \multicolumn{1}{c|}{\begin{tabular}[c]{@{}c@{}}48\%\\ (36/75)\end{tabular}}           & \multicolumn{1}{c|}{\begin{tabular}[c]{@{}c@{}}52\%\\ (39/75)\end{tabular}}           & \multicolumn{1}{c|}{\begin{tabular}[c]{@{}c@{}}-12 s\\ (-38\%)\end{tabular}}       & \begin{tabular}[c]{@{}c@{}}-15\\ min\end{tabular}                  & \multicolumn{1}{c|}{\textbf{\begin{tabular}[c]{@{}c@{}}93\%\\ (70/75)\end{tabular}}} & \textbf{\begin{tabular}[c]{@{}c@{}}0\% \\ (0/75)\end{tabular}}  & \multicolumn{1}{c|}{\textbf{\begin{tabular}[c]{@{}c@{}}99\%\\ (74/75)\end{tabular}}} & \textbf{\begin{tabular}[c]{@{}c@{}}0\% \\ (0/75)\end{tabular}}  \\ \hline
\multicolumn{1}{|l|}{One-shot}                                                                    & \begin{tabular}[c]{@{}c@{}}47\%\\ (94/202)\end{tabular}           & \multicolumn{1}{c|}{\begin{tabular}[c]{@{}c@{}}75\%\\ (68/91)\end{tabular}}            & \multicolumn{1}{c|}{\begin{tabular}[c]{@{}c@{}}21\%\\ (19/91)\end{tabular}}           & \multicolumn{1}{c|}{\begin{tabular}[c]{@{}c@{}}145 MB\\ (25\%)\end{tabular}}          & 13 GB                                                         & \multicolumn{1}{c|}{\begin{tabular}[c]{@{}c@{}}59\%\\ (54/91)\end{tabular}}           & \multicolumn{1}{c|}{\begin{tabular}[c]{@{}c@{}}41\%\\ (37/91)\end{tabular}}           & \multicolumn{1}{c|}{\textbf{\begin{tabular}[c]{@{}c@{}}32 s\\ (6\%)\end{tabular}}} & 49 min                                                             & \multicolumn{1}{c|}{\begin{tabular}[c]{@{}c@{}}87\%\\ (79/91)\end{tabular}}          & \textbf{\begin{tabular}[c]{@{}c@{}}0\% \\ (0/91)\end{tabular}}  & \multicolumn{1}{c|}{\begin{tabular}[c]{@{}c@{}}97\%\\ (88/91)\end{tabular}}           & \textbf{\begin{tabular}[c]{@{}c@{}}0\% \\ (0/91)\end{tabular}}  \\ \hline
\multicolumn{1}{|l|}{20-shot}                                                                     & \begin{tabular}[c]{@{}c@{}}49\%\\ (99/202)\end{tabular}           & \multicolumn{1}{c|}{\begin{tabular}[c]{@{}c@{}}73\%\\ (72/99)\end{tabular}}            & \multicolumn{1}{c|}{\begin{tabular}[c]{@{}c@{}}22\%\\ (22/99)\end{tabular}}           & \multicolumn{1}{c|}{\begin{tabular}[c]{@{}c@{}}245 MB\\ (18\%)\end{tabular}}         & 24 GB                                                         & \multicolumn{1}{c|}{\begin{tabular}[c]{@{}c@{}}58\%\\ (57/99)\end{tabular}}           & \multicolumn{1}{c|}{\begin{tabular}[c]{@{}c@{}}42\%\\ (42/99)\end{tabular}}           & \multicolumn{1}{c|}{\begin{tabular}[c]{@{}c@{}}30 s\\ (-5\%)\end{tabular}}         & 49 min                                                             & \multicolumn{1}{c|}{\begin{tabular}[c]{@{}c@{}}80\%\\ (79/99)\end{tabular}}          & \textbf{\begin{tabular}[c]{@{}c@{}}0\% \\ (0/99)\end{tabular}}  & \multicolumn{1}{c|}{\begin{tabular}[c]{@{}c@{}}96\%\\ (95/99)\end{tabular}}          & \textbf{\begin{tabular}[c]{@{}c@{}}0\% \\ (0/99)\end{tabular}}  \\ \hline
\multicolumn{1}{|l|}{30-shot}                                                                     & \begin{tabular}[c]{@{}c@{}}54\%\\ (108/202)\end{tabular}          & \multicolumn{1}{c|}{\begin{tabular}[c]{@{}c@{}}81\%\\ (88/108)\end{tabular}}           & \multicolumn{1}{c|}{\begin{tabular}[c]{@{}c@{}}15\%\\ (16/108)\end{tabular}}          & \multicolumn{1}{c|}{\begin{tabular}[c]{@{}c@{}}274 MB\\ (25\%)\end{tabular}}          & 29 GB                                                         & \multicolumn{1}{c|}{\begin{tabular}[c]{@{}c@{}}59\%\\ (64/108)\end{tabular}}          & \multicolumn{1}{c|}{\begin{tabular}[c]{@{}c@{}}41\%\\ (44/108)\end{tabular}}          & \multicolumn{1}{c|}{\begin{tabular}[c]{@{}c@{}}22 s\\ (-1\%)\end{tabular}}         & 39 min                                                             & \multicolumn{1}{c|}{\begin{tabular}[c]{@{}c@{}}80\%\\ (86/108)\end{tabular}}         & \textbf{\begin{tabular}[c]{@{}c@{}}0\% \\ (0/108)\end{tabular}} & \multicolumn{1}{c|}{\begin{tabular}[c]{@{}c@{}}95\%\\ (103/108)\end{tabular}}        & \textbf{\begin{tabular}[c]{@{}c@{}}0\% \\ (0/108)\end{tabular}} \\ \hline
\multicolumn{1}{|l|}{50-shot}                                                                     & \textbf{\begin{tabular}[c]{@{}c@{}}63\%\\ (128/202)\end{tabular}} & \multicolumn{1}{c|}{\textbf{\begin{tabular}[c]{@{}c@{}}82\%\\ (105/128)\end{tabular}}} & \multicolumn{1}{c|}{\textbf{\begin{tabular}[c]{@{}c@{}}13\%\\ (16/128)\end{tabular}}} & \multicolumn{1}{c|}{\textbf{\begin{tabular}[c]{@{}c@{}}322 MB\\ (32\%)\end{tabular}}} & \textbf{35 GB}                                                & \multicolumn{1}{c|}{\textbf{\begin{tabular}[c]{@{}c@{}}70\%\\ (89/128)\end{tabular}}} & \multicolumn{1}{c|}{\textbf{\begin{tabular}[c]{@{}c@{}}30\%\\ (39/128)\end{tabular}}} & \multicolumn{1}{c|}{\begin{tabular}[c]{@{}c@{}}24 s\\ (6\%)\end{tabular}}          & \textbf{52 min}                                                    & \multicolumn{1}{c|}{\begin{tabular}[c]{@{}c@{}}77\%\\ (98/128)\end{tabular}}         & \begin{tabular}[c]{@{}c@{}}2\% \\ (2/128)\end{tabular}          & \multicolumn{1}{c|}{\begin{tabular}[c]{@{}c@{}}91\%\\ (117/128)\end{tabular}}        & \begin{tabular}[c]{@{}c@{}}1\% \\ (1/128)\end{tabular}          \\ \hline
\multicolumn{1}{|l|}{Developer}                                                                   & \begin{tabular}[c]{@{}c@{}}47\%\\ (94/202)\end{tabular}           & \multicolumn{1}{c|}{\begin{tabular}[c]{@{}c@{}}53\%\\ (50/94)\end{tabular}}            & \multicolumn{1}{c|}{\begin{tabular}[c]{@{}c@{}}35\%\\ (33/94)\end{tabular}}           & \multicolumn{1}{c|}{\begin{tabular}[c]{@{}c@{}}79 MB\\ (4\%)\end{tabular}}            & 7 GB                                                          & \multicolumn{1}{c|}{\begin{tabular}[c]{@{}c@{}}52\%\\ (49/94)\end{tabular}}           & \multicolumn{1}{c|}{\begin{tabular}[c]{@{}c@{}}48\%\\ (45/94)\end{tabular}}           & \multicolumn{1}{c|}{\begin{tabular}[c]{@{}c@{}}-2 s\\ (-41\%)\end{tabular}}        & -4 min                                                             & \multicolumn{1}{c|}{\begin{tabular}[c]{@{}c@{}}56\%\\ (53/94)\end{tabular}}          & \begin{tabular}[c]{@{}c@{}}7\% \\ (7/94)\end{tabular}           & \multicolumn{1}{c|}{\begin{tabular}[c]{@{}c@{}}82\%\\ (77/94)\end{tabular}}          & \begin{tabular}[c]{@{}c@{}}6\% \\ (6/94)\end{tabular}           \\ \hline

\end{tabular}

\label{table:prompting_techniques}
\end{adjustbox}
\end{table*}

We build these Dockerfiles at every commit in the project's history, covering 1,519 Dockerfile commits. At each commit, we measure the image size and build duration. We note that in instances of build failures, which occurred in 43\% (653/1,519) of the cases, we maintain continuity by using the image size and build duration from the next successful commit. In rare cases where the Dockerfile failed to build on the last commit (one case), we use the previous successful commit. We identify three outliers with image size increases of 300-500\% between two commits due to entire Dockerfile replacements. After removing these outliers, \textit{our analysis focuses on 72 Dockerfiles, covering a total of 1,459 Dockerfile commits}. \autoref{fig:build_image} illustrates the mean increase in Dockerfile image size and build duration relative to the initial commit across project lifecycle stages. 

\textbf{Dockerfile image size grows initially, peaks mid-project, and then stabilizes.} The early stages (1-3) of the project show a steady increase in Docker image size, with an average growth rate of 15\% by stage 3. This upward trend continues into the middle stages (4-6), peaking at stage 6 with a 31\% increase.  In the later stages (7-10), the image size decreases slightly and stabilizes around a 30\% increase from the original size. 

\textbf{Dockerfile build duration rises steadily, peaking at the project’s end.} Initially (1-3), the build duration increases, reaching around 20\% by stage 3. In the mid-stages (4-6) there is a  notable increase, with build duration rising to about  41\%. In the later stages (7-10), this trend peaks at around 50\% by stage 8 and then remains relatively constant towards the end.  
\vspace{-5pt}
\begin{findingbox}
\textbf{\faKey\ Finding-1.} Docker projects in our dataset shows a mean increase of 30\% in image size and 50\% in build duration from initial development to project maturity.
\end{findingbox} 
\vspace{-5pt}
In addition to analyzing the evolution of performance metrics, we investigate the temporal patterns of Dockerfile refactoring activities across Dockerfiles commits history. Out of 1,459 commits, 739 involve refactoring actions.

\textbf{Most developers start refactoring Dockerfiles in the mid to later stages of project development.} \autoref{fig:per_ref} shows the cumulative percentage of Dockerfiles with first refactoring action across lifecycle stages: In stages (1-3), only 40\% (28/72) had initiated refactoring. A significant uptick is observed in middle stage stages (4-7), reaching 89\% (64/72) by stage 7.  

\textbf{Refactoring activities are most prevalent during the middle stages of project development, where they constitute more than half of the Dockerfile commits}. \autoref{fig:propo_ref} shows the mean proportion of Dockerfile refactoring commits in relation to all Dockerfile commits at each stage of the lifecycle, highlighting the frequency of refactorings during different stages. In the early stages (1-3), 
refactoring commits compromise approximately 35\% of commits. As the project advances (4–7), there is a notable increase in refactoring, peaking at 58\% in stage 5 and remaining above 50\%.
In the later stages (8-10), the refactoring commits rise to 60\% at stage 8, before stabilizing around 45\%.
  \begin{findingbox}
\textbf{\faKey\ Finding-2.}
60\% percent of Dockerfiles in our dataset underwent their first refactoring in the mid to later stages of development. During this period, refactoring actions made up over 50\% of all commits. 
\end{findingbox}  
The interplay between Dockerfile refactoring activities and the evolution of image size and build duration reveals insights into Docker project development practices. Our analysis suggests that early project phases often experience rapid increases in Docker image size, which may be driven by a focus on feature expansion and the incorporation of dependencies, with less emphasis on refactoring. This could result in the accumulation of technical debts. As projects mature, particularly during the middle stages, an uptick in refactoring activities coincides with a peak in image size growth, potentially indicating a critical shift toward optimization. Interestingly, while these refactoring efforts may contribute to the stabilization of image size, they can also coincide with an increase in build duration, suggesting that improvements in image size often entail more complex build procedures. These observations highlight the potential benefits of initiating refactoring efforts early in the project lifecycle and maintaining them consistently to prevent the accumulation of technical debts that could complicate Dockerfile development.

\subsection{RQ2: Effectiveness of ICL in automating Dockerfile refactoring using zero-shot, one-shot, and few-shot approaches} 

In RQ2, we investigate the use of LLMs to automate the process of Dockerfile refactoring. Our study involved different configurations, specifically zero-shot, one-shot, 20-shot, 30-shot, and 50-shot approaches. The results, detailed in \autoref{table:prompting_techniques}, emphasize the effectiveness of LLM and the impact of the number of demonstrations on the rate of successful builds and the reduction of technical debts.



\textbf{The build success rate exhibits a positive correlation with the number of refactoring demonstrations provided.} In zero-shot, only 38\% (77/202) of Dockerfiles were successfully built. This success rate increased to 47\% (94/202) in one-shot, 49\% (99/202) in 20-shot,  54\% (108/202) in 30-shot, and reached the highest at 63\% (128/202) in 50-shot. 

We conducted a behavior assessment for Dockerfiles that were successfully built in all settings. We also evaluate key metrics—image size, build duration, understandability, and maintainability—using improvement rate, deterioration rate, and average reduction. The improvement rate measures the proportion of Dockerfiles showing positive changes post-refactoring, while the deterioration rate indicates the proportion experiencing negative changes. Average reduction quantifies the magnitude of improvement among all the successfully built Dockerfiles. 
The behavior assessment revealed a very low incidence of functional changes, with only 5 cases (2 zero-shot, 3 one-shot) where Dockerfile behavior changed post-refactoring. These were excluded from our calculations.

\textbf{The average reduction rate in image size reaches its highest values with more refactoring demonstrations.}When examining image size, the improvement rate was consistently high across most settings, hovering around 81\% for zero-shot, one-shot, and 50-shot and 72\% and 75\%  respectively for the one-shot and 20-shot. Regarding average image size reduction, zero-shot achieved an average of 18\% (129 MB), whereas both one-shot and 30-shot settings achieved similar reduction rates of around 25\%. The 50-shot setting showed the most significant reduction, with an average of 32\% (322 MB), resulting in a total saving of 35 GB across all Dockerfiles.

\textbf{The Build duration average reduction fluctuates, showing the highest improvement with more refactoring demonstrations.} Upon analyzing the build duration, we find that the zero-shot setting has the lowest rate of improvement at 48\%. This suggests that over half of the successful builds experienced longer build duration after the refactoring. The one-shot, 20-shot, and 30-shot configurations exhibit similar enhancement, achieving improvement rates of approximately 58-59\%. The 50-shot configuration demonstrates the most notable improvement rate, reaching 70\%. In terms of the average reduction in build duration, the one-shot and 50-shot settings were the most successful, resulting in average reductions of 6\% (32 seconds and 24 seconds, respectively). The total amount of time saved using the 50-shot setting was 52 minutes.


\textbf{Understandability shows an inverse correlation with the number of refactoring demonstrations.} In the zero-shot setting, the understandability improvement rate is 93\%, but this rate decreased to 87\% in the one-shot setting and further to 77\% in the 50-shot setting.

\textbf{Maintainability shows an inverse correlation with the number of refactoring demonstrations provided, with high overall results across all settings.} Maintainability was highest in the zero-shot setting at a rate of 99\%. This slightly decreased to 95\% in the 30-shot setting and 91\% in the 50-shot setting. Despite these reductions, maintainability remained high across all settings.
\vspace{-20pt}
\begin{findingbox}
\textbf{\faKey\ Finding-3.} 
Increasing refactoring demonstrations to 50-shot improves ICL performance in build success rate (63\%), image size reduction (32\%), and build duration reduction (6\%). However, this negatively affects maintainability and understandability, though the improvement rates remain high at 77\% and 91\%, respectively.
\end{findingbox}
\vspace{-6pt}
Furthermore, we study the correlation between different key metrics during the 50-shot setting to understand whether refactorings improve these metrics together or if the improvement of one metric leads to the deterioration of another. 
Using Spearman correlation \cite{spearman1961proof}, we find a moderate negative correlation (-0.60) between image size and build duration, with smaller images often increasing build duration. A strong negative correlation (-0.80) between build duration and both understandability and maintainability suggests shorter builds may reduce clarity and maintainability. Understandability and maintainability are perfectly correlated (1.00), while image size has no effect on them (0.00).The correlation matrix can be found in \cite{ICSE25_replication}.
\vspace{-20pt}
\begin{findingbox}
\textbf{\faKey\ Finding-4.} 
Improvements in image size often lead to longer build durations, and reducing build duration tends to decrease understandability and maintainability. Conversely, improvements in understandability directly enhance maintainability.
 \end{findingbox}
 \vspace{-5pt}

\begin{figure}[!b]
    \centering
    \includegraphics[scale=0.19]{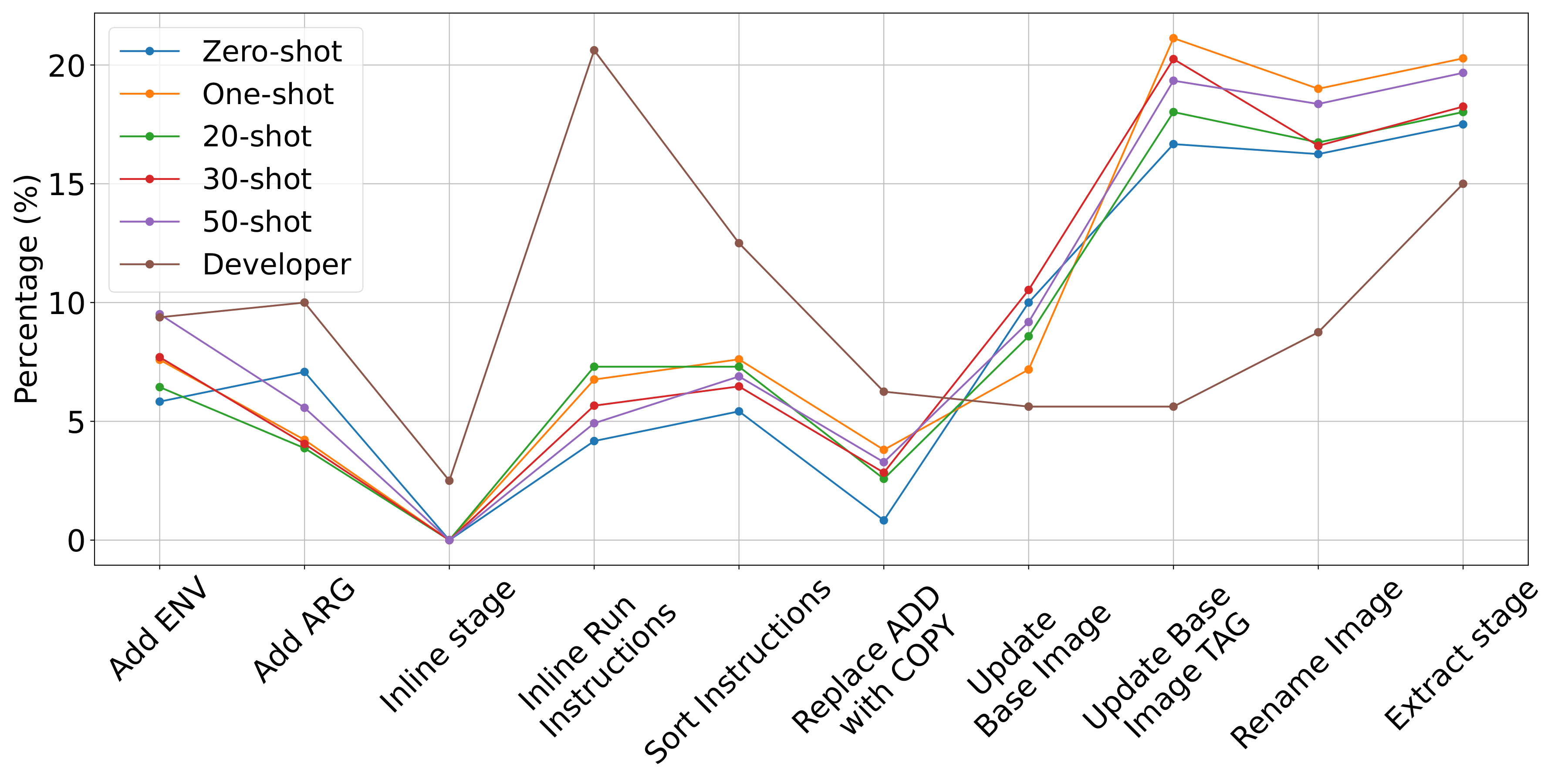}
    \caption{Refactoring Techniques Distribution}
    \label{fig:corro} 
\end{figure}

\textbf{The distribution of refactoring techniques is nearly consistent across all prompting methods}, as illustrated in \autoref{fig:corro}. This consistency is observed in zero-shot, one-shot, and few-shot settings, where techniques such as ``Extract Stage,'' ``Rename Image'', and ``Update Image Tag'' are notably prevalent, each with frequencies exceeding 15\%. These techniques emphasize enhancing image size, maintainability (version control), and understandability (renaming). On the other hand, techniques such as ``Inline Run Instruction'' and ``Sort Instruction'' consistently exhibit lower frequencies, generally around 4\% - 7\%. Notably, the ``Inline Stage'' refactoring was not performed in any setting. 

\begin{findingbox}
\textbf{\faKey\ Finding-5.} 
Regardless of the number of demonstrations, the LLM consistently applied the same refactoring techniques. However, varied evaluation metrics indicate that the specific implementation and choices made (e.g., base image, tag, stages, etc.) during the refactoring process impact its effectiveness.
\end{findingbox}
\vspace{-2mm}

\subsection{RQ3: Effectiveness of automated Dockerfile refactoring in improving quality compared to manual refactoring and smell-detection tools.}


While metrics show significant improvements through automation of Dockerfile refactoring, comparing these results with human developers' work is essential for practical validation. We benchmark against human refactorings using the 50-shot setting, our best performer, to identify strengths and areas for improvement.

We have a manually refactored version of the 202 test Dockerfiles. We compute key metrics for these developer versions, such as build success rate, image size, build duration, maintainability, and understandability, as shown in \autoref{table:prompting_techniques}. Developers achieved a build success rate of 47\% (92/202), while the 50-shot setting achieved 63\% (128/202). 

To ensure a detailed comparison between automated and manual refactoring, we focused on the 66 Dockerfiles that were successfully built by both methods and analyzed improvements in build duration, image size, maintainability, and understandability for this intersection set.

\begin{figure}[!t]
    \centering
    \includegraphics[width=1\linewidth]{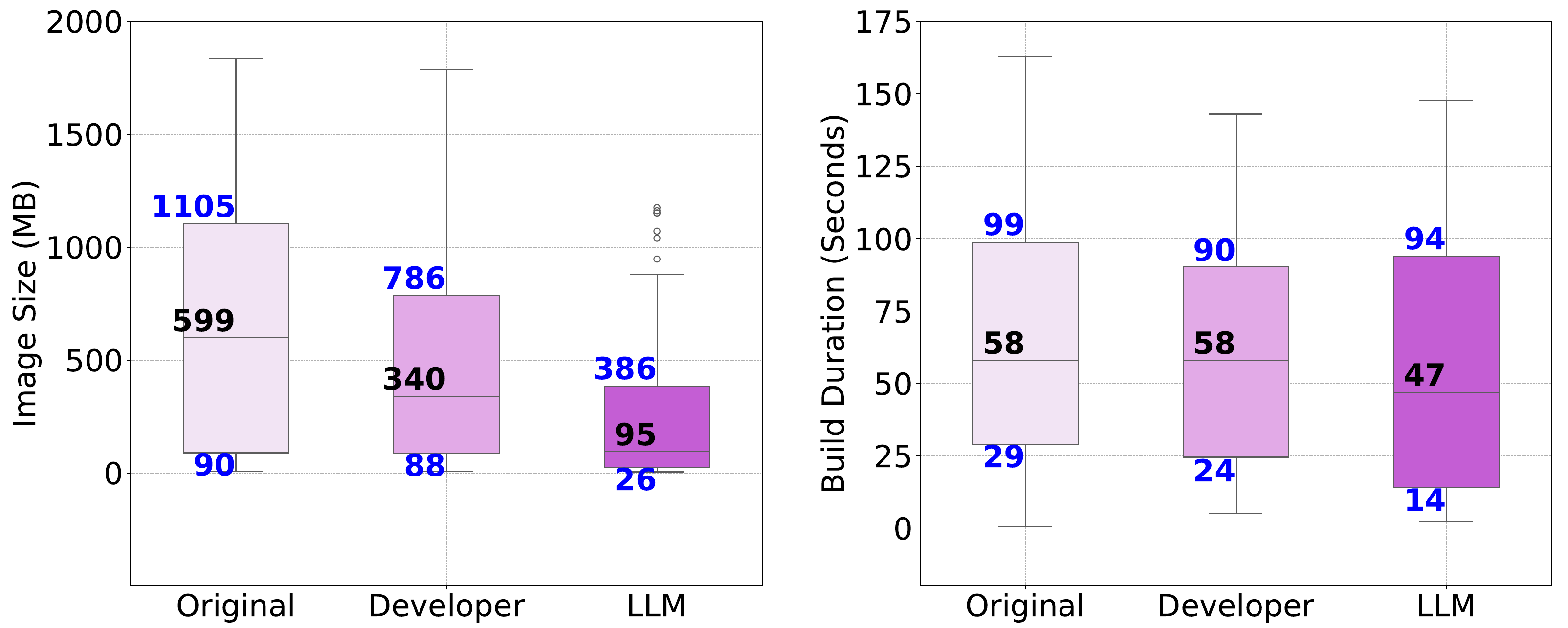}
    \caption{ Docker image sizes and build durations across three categories: the original Dockerfiles (pre-refactoring), those refactored by developers, and those refactored using the 50-shot LLM approach.}
    \label{fig:double_box} 
\end{figure}

\textbf{Automated refactoring outperformed developer refactoring, reducing median image size twice as effectively} As illustrated in \autoref{fig:double_box}, the original Dockerfiles have a median image size of 599 MB. Developer refactoring reduces this to 340MB, a 43\% reduction. The 50-shot LLM refactoring achieves a more substantial reduction, bringing the median size down to 95 MB, an 85\% reduction.

\textbf{Automated refactoring achieves a much greater reduction in build duration compared to developer refactoring.} As illustrated in \autoref{fig:double_box}, the original Dockerfiles exhibit a median build duration of 58 seconds. The developer's refactoring maintained the median build duration at 58 seconds, indicating that the central tendency of the build duration did not improve. The 50-shot LLM refactoring achieves a more significant reduction, bringing the median duration down to 46 seconds, a 19\% reduction.

\textbf{Automated refactoring shows a higher understandability improvement rate compared to developer refactorings} For understandability, LLM refactoring improved 80\% of cases (53/66), maintained 18\% (12/66), and worsened one case. Developer improvement was 57\% (38/66), maintained 38\% (25/66), and worsened in three cases.

\textbf{Automated refactoring shows a higher maintainability improvement rate compared to developer refactoring, with strong results for both}. For maintainability, LLM refactoring improved 91\% of cases (60/66), maintained 7\% (5/66), and worsened one case. Developer improvement was 71\% (47/66), maintained 21\% (14/66), and worsened in five cases.
 \vspace{-5pt}
\begin{findingbox}
\textbf{\faKey\ Finding-6.} 
Automated refactoring outperformed developers' manual refactorings on all metrics, reducing image size twice as much and achieving around 19-23\% better improvements in build duration, understandability, and maintainability. \end{findingbox} 
\vspace{-5pt}

\begin{figure}[!t]
    \centering
    \includegraphics[scale=0.32]{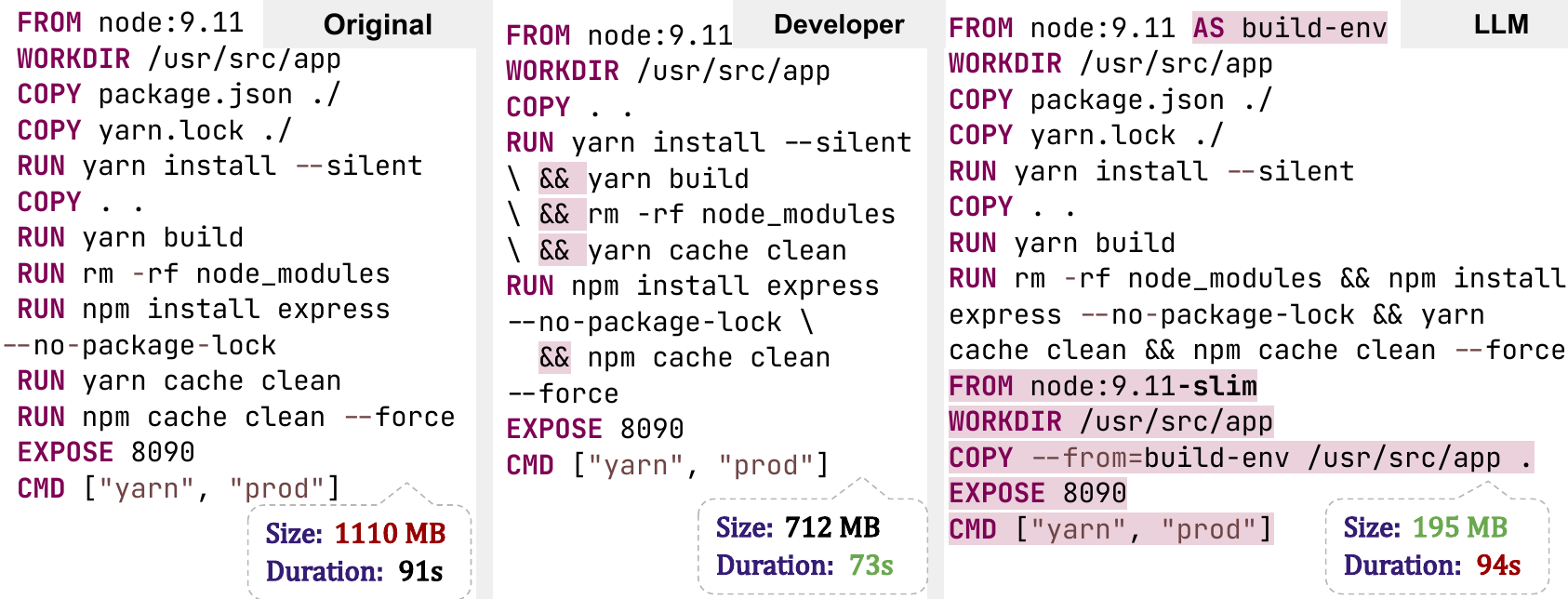}
    \caption{Illustrative Example of Dockerfile Refactoring: Manual vs. Automated (Commit 729ee76 from cars10/elasticvue; comments have been removed)}
    \label{fig:example}
\end{figure}

\textbf{Developers and LLM exhibit different distributions of refactoring techniques, with the exception of the ``Extract Stage'' technique, which is frequently used by both.} The plot in \autoref{fig:example} shows distinct patterns in refactoring techniques between developers and LLMs. Interestingly, both groups use the ``Extract Stage'' technique frequently, accounting for around 15\% of total refactorings, highlighting a shared emphasis on this method. Additionally, developers predominantly use the ``Inline Run Instruction'' technique, representing 20\% of their refactorings. In contrast, techniques like ``Update Image Tag'', ``Inline Stage'', and ``Update Base Image'' are less frequently used by developers, suggesting a lesser focus on these methods.

\begin{table}[!b]
\caption{Refactoring vs. Smell Fixing: Image Size Reduction and Build Success Rate}
\label{tab:comparison}
\centering
\begin{adjustbox}{max width=0.5\textwidth} 
\begin{tabular}{|l|c|c|c|c|c|}
\hline
\multicolumn{1}{|l|}{\begin{tabular}[c]{@{}l@{}}Tool / \\ Approach\end{tabular}} & 
\begin{tabular}[c]{@{}c@{}}Build \\ Success \\ Rate (\%)\end{tabular}  & 
\begin{tabular}[c]{@{}c@{}}Improvement \\ Rate (\%)\end{tabular}      & 
\begin{tabular}[c]{@{}c@{}}Deterioration \\ Rate (\%)\end{tabular}   & 
\begin{tabular}[c]{@{}c@{}}Average \\ Reduction \\ (MB)\end{tabular}            & 
\begin{tabular}[c]{@{}c@{}}Total \\ Reduction \\ (GB)\end{tabular} \\ \hline

Smell-Repair (PARFUM\cite{durieux2023parfum}) & \begin{tabular}[c]{@{}c@{}}50.4\% \\ (63/125)\end{tabular} & 
\begin{tabular}[c]{@{}c@{}}69.8\% \\ (44/63)\end{tabular} & 
\begin{tabular}[c]{@{}c@{}}30.2\% \\ (19/63)\end{tabular} & 
\begin{tabular}[c]{@{}c@{}}12.2 MB \\ (2.2\%)\end{tabular} & 
\begin{tabular}[c]{@{}c@{}}0.8 GB\end{tabular} \\ \hline

Refactoring (ICL) & \begin{tabular}[c]{@{}c@{}}66.4\% \\ (83/125)\end{tabular} & 
\begin{tabular}[c]{@{}c@{}}88.0\% \\ (73/83)\end{tabular} & 
\begin{tabular}[c]{@{}c@{}}12.0\% \\ (10/83)\end{tabular} & 
\begin{tabular}[c]{@{}c@{}}212.0 MB \\ (25.6)\%\end{tabular} & 
\begin{tabular}[c]{@{}c@{}}17.1 GB\end{tabular} \\ \hline

\end{tabular}
\end{adjustbox}
\end{table}
\underline{\textit{Case analysis:}} In \autoref{fig:example}, we present a case to demonstrate the difference in Dockerfile refactoring strategies between the developer and the LLM.
The original Dockerfile (pre-refactoring), using a node:9.11 base image, resulted in an image size of 1110 MB and a build duration of 91 seconds. This Dockerfile included multiple steps such as copying dependency files, installing packages, building the application, and cleaning up unnecessary files. The human developer's refactoring introduced significant improvements. The image size was reduced to 712 MB, and the build duration decreased to 73 seconds. This was achieved through ``Inline Run Instruction'' refactoring, which involved grouping commands in a single \textit{RUN} instruction using \&\&. This technique minimizes the number of Docker layers created. Additionally, the developer removed the redundancy of two \textit{COPY} commands, initially intended to optimize caching for dependency installation, by consolidating them into a single step. This further contributed to the reduced image size. In contrast, the LLM's approach demonstrated a more complex strategy. The LLM employed a multi-stage build process, often referred to as ``Extract Stage'' refactoring. This approach ensured that only essential components for running the application were included in the final image, while all build dependencies were discarded. Furthermore, the LLM transitioned from a node:9.11 to a node:9.11-slim image for the final stage, a technique known as ``Update Image Tag''. These actions significantly reduced the image size to 195 MB, although the build duration slightly increased to 94 seconds. The original Dockerfiles and those refactored by the developers and the language model all have similar levels of maintainability and understandability. This consistency is due to avoiding unspecified versions, which supports maintainability, and including detailed comments for each step, which enhances understandability.

Alongside the comparisons of automated and manual refactoring, we assessed the reduction in image size achieved through automated refactoring in contrast to the automated smell repair performed by PARFUM on the smelly Dockerfiles present in our dataset (125 Dockerfiles and 74 projects).

\textbf{Automated refactoring achieves a 10x reduction in Dockerfile image size and a higher build success rate compared to automated smell repair}. As illustrated in \autoref{tab:comparison}, ICL refactoring achieved a build success rate of 66.4\% (83/125), notably higher than PARFUM's 50.4\% (63/125). Moreover, automated refactoring yielded an improvement rate of 88.0\%, surpassing PARFUM's improvement rate of 69.8\%. Notably, the average image size reduction achieved by ICL was 212.0 MB (25.6\%), translating to a total reduction of 17.1 GB across all Dockerfiles. In comparison, PARFUM attained an average reduction of 12.2 MB (2.2\%) and a total reduction of only 0.8 GB.

These findings reveal a key limitation of smell-focused approaches. While PARFUM offers modest image size gains and outperforms other smell-fixing tools \cite{durieux2024empirical}, it remains confined to fixing specific issues within individual instructions, mainly addressing Bash commands in RUN instructions. By focusing narrowly on tasks like cache clearing, temporary file removal, and package installation cleanup (e.g., pip, npm, apt-get), PARFUM achieves limited image size reduction and fails to address structural inefficiencies in Dockerfiles. In contrast, refactoring adopts a design-level approach by reconfiguring all instructions and consolidating layers across stages.

\subsection{RQ4: Build Failure Reasons}

Despite notable improvements in various metrics, build failure rates remain relatively high for both manual and automated refactoring, at 53\% (108/202) and 37\% (74/202) respectively.

Ensuring Dockerfile refactorings result in functional builds is imperative; otherwise, the refactoring process defeats its purpose by causing build failures. A qualitative analysis of these failures reveals their primary causes. 


\textbf{Build context errors} were the most frequent cause of failures in LLM-generated Dockerfiles, accounting for 52\% (39/74) of the cases, compared to 33\% (36/108) in developer-refactored Dockerfiles. Notably, there were 27 instances of build context errors common to both LLM-generated and developer-refactored Dockerfiles. These errors often arise from incorrect file paths or misconfigured build contexts. In this example \cite{ICSE25_replication_Example1} , the error occurred because the path to ``install.sh'' in the \textit{COPY} instruction did not consider the build context. 

\textbf{Dependency errors} were the most prevalent issue in developer-refactored Dockerfiles, representing 43\% (46/108) of failures, and were also significant in LLM-generated Dockerfiles at 27\% (20/74), with 8 cases of overlap. These errors typically occur when software components, packages, libraries, or tools are missing, incompatible, or incorrectly configured. An example is found in \cite{ICSE25_replication_Example2}, where developers performed an ``Update Image Tag'' to an older Debian ``wheezy'' base image, causing compatibility issues and ``apt-get install'' failures due to outdated repositories. 



\textbf{Syntax errors} were more common in LLM-generated Dockerfiles, constituting 11\% (8/74) of failures, compared to 6\% (7/108) in developer-refactored ones, indicating a need for enhanced syntax validation in LLM processes. For instance, in one example, \cite{ICSE25_replication_Example3}, the ``\&\&'' was missing when performing ``Inline Run Instruction'' refactoring. 


\textbf{Missing base images} were more frequently an issue in developer-refactored Dockerfiles, occurring in 15\% (16/108) of cases. This was often due to developers using local images not available on DockerHub, leading to build failures due to the inability to retrieve these private images. Conversely, the rate for LLM-generated Dockerfiles was 5\% (4/74), primarily due to the generation of references to non-existent images. For example, a Dockerfile failed in \cite{ICSE25_replication_Example4} because developers performed an ``Update Base Image'' refactoring and specified a base image that was not present in Dockerhub. 

Other errors, related to experimental settings such as resource limitations and network issues, were very less common.

\


\section{Threats to validity}
\label{sec:07-threats}

This study acknowledges several threats to validity.
A potential one is the diversity of our dataset. Although we collected a large set of Dockerfiles using various refactoring keywords, but our results may not fully encompass Dockerfiles that require unique build steps. During data collection, we focused on immediate builds and excluded files needing additional setup or dependencies. However, the dataset's inclusion of projects from various domains, with varying popularity and programming languages, provides a reasonably representative sample. Future research can replicate our study across different scenarios to ensure broader applicability. Additionally, there is a potential risk that some test Dockerfiles may have been encountered by GPT-4o during its pre-training phase, which could influence the results. However, the model's suboptimal zero-shot performance and superior results compared to developer refactorings suggest that memorization is unlikely to be the primary driver of outcomes. The metrics used, such as image size and build duration, can be affected by factors beyond Dockerfile modification, such as application changes and Docker engine conditions. We mitigated this by standardizing testing conditions, including clearing Docker caches and using consistent commit points. We relied on DRMiner \cite{ksontini2024drminer}, a state-of-the-art tool, to detect refactorings between commits. Despite an F1 score of 0.94, DRMiner’s limitations may lead to missed or misidentified refactorings.

\section{Conclusion}
\label{sec:conclusion}
Throughout the lifecycle of Docker projects, we observe that image size and build duration consistently increase. This phenomenon is often due to the gradual accumulation of dependencies and suboptimal configurations, with developers frequently postponing necessary refactorings. 

In exploring the automation of refactoring, we demonstrated that LLMs can significantly streamline this process. Our experiments revealed that the effectiveness of these models improves with the number of refactoring demonstrations provided, with the 50-shot configuration yielding the most substantial gains. Specifically, this setup achieved superior results in reducing Docker image size and build duration and also improved maintainability and understandability compared to manual refactoring by developers. A key finding of our study is that refactoring involves trade-offs; enhancing one metric often negatively impacts others. Moreover, both automated and manual refactorings are susceptible to build failure, primarily due to context errors and dependency issues. This issue emphasizes the need for more robust error handling and validation mechanisms, potentially as part of future work.

Integrating automated refactoring into CI/CD pipelines can provide a more agile and responsive development environment, allowing for continuous improvement and optimization of Dockerfiles. This study contributes to the growing body of knowledge on the application of AI in IaC.

\bibliographystyle{IEEEtran}
\bibliography{references}
\end{document}